\documentclass[aps,prl,twocolumn,superscriptaddress,groupedaddress]{revtex4}  % for review and submission
\usepackage{graphicx}  % needed for figures
\usepackage{dcolumn}   % needed for some tables
\usepackage{bm}        % for math
\usepackage{amssymb}   % for math
\usepackage{amsmath}   % for math
\usepackage{mathtools} % for math
\usepackage{color}
\usepackage{ulem}

\usepackage[colorlinks,linkcolor=blue,urlcolor=blue,citecolor=blue]{hyperref}
 
\begin{document}

\title{Different anomalous diffusion regimes measured in the dynamics of tracer particles in actin networks}
\author{Maayan Levin}
\affiliation{School of Chemistry, Raymond and Beverly Sackler Faculty of Exact Sciences, Tel Aviv University, Tel Aviv 6997801, Israel}
\affiliation{Center for the Physics and Chemistry of Living Systems. Tel Aviv University, 6997801, Tel Aviv, Israel}
\author{Golan Bel}
\affiliation{Department of Solar Energy and Environmental Physics, Blaustein Institutes for Desert Research, Ben-Gurion University of the Negev, Sede Boqer Campus 8499000, Israel}
\author{Yael Roichman}
\email{roichman@tauex.tau.ac.il}
\affiliation{School of Chemistry, Raymond and Beverly Sackler Faculty of Exact Sciences, Tel Aviv University, Tel Aviv 6997801, Israel}\affiliation{School of Physics \& Astronomy, Raymond and Beverly Sackler Faculty of Exact Sciences, Tel Aviv University, Tel Aviv 6997801, Israel}
\affiliation{Center for the Physics and Chemistry of Living Systems. Tel Aviv University, 6997801, Tel Aviv, Israel}

\date{\today}

\begin{abstract}
\noindent
It was previously believed that diffusion of a tracer particle in a viscoelastic material should be of the fractional Brownian motion (fBm) type. This is due to the long-term memory in the response of such materials to mechanical perturbations. Surprisingly, the diffusion of a tracer particle in a network of a purified protein, actin, was found to conform to the continuous time random walk (CTRW) type in one study that focused on the ensemble average characteristics. Here, we analyze the dynamics of two differently sized tracer particles in actin networks of different mesh sizes. We find that the ratio of tracer particle size to the characteristic length scale of a bio-polymer network plays a crucial role in determining the type of diffusion the particle performs. We find that the tracer particle diffusion has features of fBm when the particle is large compared to the mesh size, of normal diffusion when the particle is much smaller than the mesh size, and of CTRW in between these two limits. Based on our findings, we propose and verify numerically a new model for the tracer particle’s motion in all regimes. Our model suggests that diffusion in actin networks consists of fBm of the tracer particle coupled with occasional caging events with power-law-distributed escape times.

\end{abstract}

\maketitle
\section{Introduction}
The proliferation of fluorescence single particle microscopy in the past decades has provided bountiful data on single entities undergoing stochastic motion in biological and bio-inspired materials \cite{Mark2017,3D,Pang2011,Bo2020,toolbox,Natalie}.
The goal of these experiments, and the accompanying theoretical works, is to learn about the characteristics of the medium in which the motion takes place and about the dynamics of the tracer particles in the usually crowded and disordered medium.
%medium info on polymer netwroks
These data reflect multiple aspects of the biological systems, such as the processes governing their fluctuations in space, their structure, and their mechanical properties.

 Successful toolkits for analyzing tracer motion have been developed in the context of random processes leading to anomalous diffusion \cite{toolbox,tools2,tools3}. In this context, the diffusion of a tracer particle is characterized by the diffusion exponent, $\alpha$. This exponent is defined by the time dependence of the mean squared displacement (MSD), $\text{MSD}\sim\tau^\alpha$, where $\tau$ is the duration of the tracer motion. We note that a concentrated effort to provide tools for efficient and accurate characterization of diffusion type in more realistically constrained conditions is underway. New approaches for data analysis include, for example, machine learning \cite{Granik2019,Bo2019,Munoz-Gil2020,Kowalek2019,Vangara2020}, Bayesian statistics \cite{Thapa2018}, and power spectrum analysis \cite{Krapf2018}.

Generally speaking, in thermal equilibrium, the diffusion of a tracer particle in a dense, complex medium is either normal or sub-diffusive. Namely, $\alpha$ is equal to or lower than 1, respectively. Sub-diffusion was observed in a variety of complex systems, including, for example, biological systems such as living cells \cite{living,living2,living3},  liquid crystals \cite{crys,crys2}, and geological systems \cite{geolo,geolo3}. Very different mechanisms can induce a sub-diffusive behavior (i.e., $\alpha<1$) including confinement, transient trapping, and memory in the mechanical response of the environment \cite{sub0,sub1,sub2,sub3,toolbox}. These mechanisms lead to different classes of diffusion processes, such as random walk on a fractal structure (RWF) \cite{RWF1,RWF2,RWF3}, continuous time random walk (CTRW) \cite{RW_book,CTRW1,CTRW2}, and fractional Brownian motion (fBm) \cite{fBM1,fBM2,fBM3}.
%Several analysis tools were developed over the years to differentiate between the various classes of sub-diffusion starting from single particle trajectories and are summarized in \cite{Meroz2015,tools2,tools3}.

In order to validate, adapt, and improve the available analytical tools, especially for use in biological systems, they need to be implemented on relevant well-controlled model systems. One candidate for such a model system is a material made of a network of the biopolymer actin, which is the most abundant protein in the cytoskeleton of eukaryotic cells. Actin creates networks in the cell, at many length scales and with different structures, which can be reproduced in-vitro by selecting the polymerization buffer and the protein composition \cite{gardel2003,gardel2006,gardel2008,gardel2009,pollard1999,Cooper1983,super1,Shin2004,Haviv2008,Sonn_Segev_2017}.
Actin networks have been characterized extensively, both in terms of their mechanical properties \cite{gardel2003,gardel2006,Lee2010,Palmer1999} and their structure \cite{adar2014,Adar2014_2}.  

The motion of tracer particles in actin networks has also been studied to some extent \cite{Anomalous2004,Saxton2010,Ghosh2015}. It is expected that tracer particles with a radius, $a$, much smaller than the mesh size of the network, $\xi$, will diffuse normally ($\alpha=1$). This is expected since the tracer particles can easily pass between filaments and are mainly sensitive to the local viscous environment \cite{crowd}. The main effect of the actin network is an increased viscosity, which results in a smaller diffusion coefficient. When $a\ge \xi$, the particles are expected to be confined to a finite region of the network with small movements within this region and due to the dynamics of the actin. For the case of $a\approx\xi$, the particles are expected to sub-diffuse, thereby reflecting the elasticity and structure of the network.
In a pioneering work \cite{Anomalous2004}, the motion of tracer particles in an in-vitro actin network was studied as a function of the ratio between the mesh size of the network and the tracer particles' radius. Surprisingly, the motion of the particles at intermediate values of $a/\xi$ belonged to the CTRW class, reflecting a process with trapping events in which the escape time distribution has a heavy tail. This should be contrasted with the expected fBm model that would reflect the visco-elasticity of the actin network. Unfortunately, this study only investigated the ensemble characteristics and not the single particle time averages.

Here, we attempt to shed light on the mechanisms governing the dynamics of colloidal tracer particles in entangled actin networks by performing single particle tracking experiments. We studied the tracer dynamics in these networks with various monomer concentrations and used two differently sized tracer particles. The recorded dynamics were analyzed extensively in terms of ergodic properties, probability density functions, and temporal and spatial correlations. The analysis was performed for the entire range of tracer radius to mesh size ratios, $0.26\leq a/\xi\leq 1$.
We found that as $a/\xi$ increases, the diffusion type changes from normal to CTRW-like to fBm. We propose a model explaining the range of transport modes, and show that the simulated trajectories resemble the observed ones.

\section{Materials and Methods}

\subsection{Materials}
G-actin was purified from rabbit skeletal muscle acetone
powder \cite{1971actin}, with a gel filtration step, stored on
ice in G-buffer (5 mM Tris HCl, 0.1 mM $\text{CaCl}_2$, 0.2
mM ATP, 1 mM DTT, 0.01\% $\text{NaN}_3$, pH 7.8), and used
within two weeks. G-actin concentration was determined by absorbance, using a UV/Visible spectrophotometer (Ultraspec 2100 pro, Pharmacia) in a cuvette with a 1-cm path length and an extinction coefficient of $\epsilon_{290} = 26, 460 $ $M^{-1}cm^{-1}$. Polystyrene colloids with radii of a = $0.75 $ $\mu m$ (Polysciences, Catalog No. 09719-10) and a = $0.25$ $\mu m$ (Invitrogen, lot 1173396) were incubated for 2 h before the experiment began in a 10-mg/ml bovine serum albumin (BSA) solution (Sigma) to prevent nonspecific binding of proteins to the bead surface \cite{valentine2004}.
Actin polymerization was initiated by adding G-actin in various concentrations to an F-buffer solution (5 mM Tris HCl, 1 mM $\text{MgCl}_2$, 0.05 M KCl, 200 $\mu$M EGTA, 1 mM ATP) and gently mixing. The colloidal particles were added before mixing. The actin monomer concentration varied from $C_A = 2$\textendash$27 $ $\mu M$, corresponding to a mesh size range of $\xi = 0.95$\textendash$0.28$ $ \mu m$, respectively \cite{SCHMIDT1989}.
	
\subsection{Sample preparation}
\label{sec:sample}
Samples were prepared on glass coverslips (40 mm in diameter) coated with methoxy-terminated PEG
(polyethylene glycol, Mw=5000 g/mol, Nanocs) for 2 h before the experiment began to prevent
F-actin filaments from sticking to the glass surface. The sample chamber was $\sim150$ $\mu m$ high and sealed with paraffin wax. Actin polymerization was carried out by mixing the monomers, the buffer, and the fluorescent tracer particles together. After mixing, the suspensions were placed on the sample holder, and the sample was sealed. The recording began only after polymerization ended and the network had reached steady state conditions according to \cite{Levin2020}. In order to avoid wall effects, imaging was done at a distance of approximately $60$ $\mu m$  from the glass. The resulting F-actin networks are well described as networks of semi-flexible polymers \cite{semi1,semi2,semi3,semi4,semi5}, and their mesh size, $\xi=0.3/\sqrt{c_a}$ \cite{SCHMIDT1989}, was easily controlled by controlling the initial monomer concentration $c_a$ ($c_a$ in mg/ml and $\xi$ in $\mu m$). The range of concentration values that we used corresponds to a mesh size range of $\xi = 0.28$\textendash$0.95$ $\mu m$, thereby enabling $0.79\leq a/\xi <2.68$ for the $a=0.75$ $\mu m$ beads and $0.26\leq a/\xi\leq0.89$ for the $a=0.25$ $\mu m$ beads.
	
\subsection{Fluorescence microscopy and imaging}
Imaging the motion of the tracer particles within the suspensions was done using an Olympus IX71 epi-fluorescence microscope, at $\lambda = 480$ nm with a 40x air objective for the a = $0.75$ $ \mu m$ particles, and at $\lambda = 605$ nm with a 60x oil objective for the a = $0.25$ $\mu m$ particles. We recorded the motion of approximately 250 particles in the field of view for $\sim 6$ min using a CMOS video camera (Grasshoper 3, Point Gray) at a frame rate of 30 Hz with an exposure time of 10 ms. We used data from at least $5 \cdot 10^4$ frames per experiment and performed three experiments for each set of actin concentration and bead size. Particle tracking was done by conventional video microscopy, using the protocol of Crocker and Grier \cite{Crocker1996} implemented in MATLAB software. The distribution of the number of snapshots in each trajectory, $L_{traj}$, varies significantly between the samples: the higher the actin concentration, the longer the average trajectory length. We used at least 750 trajectories in our analysis and in most cases over 2000; the analysis is based on trajectories satisfying $L_{traj}>500$.

\section{Results and discussion}

\begin{figure}[ht!]
	\includegraphics[scale=0.2]{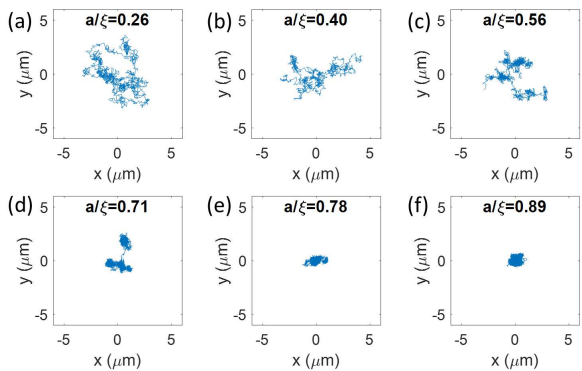}
	\caption{Typical trajectories of tracer particles projected on the x-y plane for samples with increasing a/$\xi$. In (a-b), the probe particles show an almost normal diffusion; in (c-e), they are more constrained by the network, but free to jump between different micro-environments; and in (f), the particles' movement is significantly restricted by the dense actin network.}
	\label{fig:trajectoriesVsRatio}
\end{figure}

In our experiments, we tuned the ratio between a and $\xi$ over a range of $0.26 <$ a/$\xi$ $< 1$. We then imaged the motion of the tracer particles embedded in the actin network at steady state conditions and extracted their trajectories. We observed marked differences between the typical trajectories of particles in networks of different actin concentrations (Fig.~\ref{fig:trajectoriesVsRatio}). As expected, trajectories of the same length spread further out at low actin concentrations. As the actin concentration increases, the trajectories become more compact. For low a/$\xi$, the trajectories reflect almost normal diffusion, since the probe particles can easily move through the network. At intermediate a/$\xi$, the particles are mostly constrained by the network and occasionally jump between different micro-environments (or "cages"). As the dimensions of the particles increase with respect to the mesh size, the particles become more restricted by the network, and movement between cages becomes less likely.

\begin{figure}[ht!]
	\centering
	\includegraphics[scale=0.22]{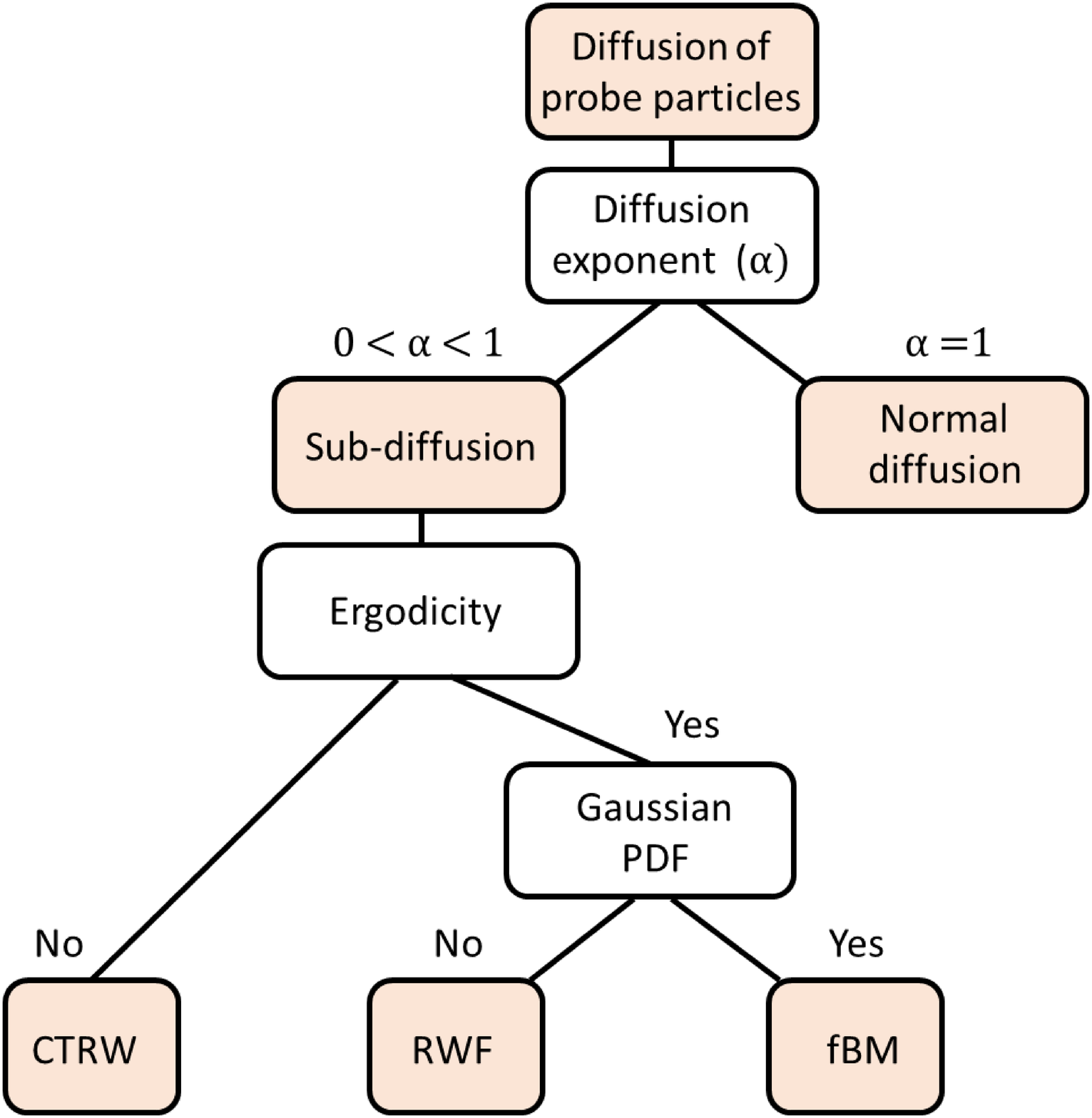}
	\caption{A decision tree for identifying the best diffusion model describing the underlying physical process of an ensemble of trajectories. Here we assume no subordinated processes. The first step includes checking whether the MSD is linear as a function of the lag time, $\tau$, i.e., $\alpha=1$, which describes normal diffusion, or whether $0 < \alpha < 1$, meaning the process is sub-diffusive. If the probe particles undergo sub-diffusive motion, we check whether the trajectories exhibit ergodicity. If they do not, then the relevant model is CTRW. If they are ergodic, we then need to find whether their probability distribution function (PDF) is Gaussian.  If it is, then the diffusion model is fBm. If it is not, then the process is best described by RWF \cite{toolbox} or other mechanisms.}
	\label{fig:flow_diagram}
\end{figure}

 In order to characterize the diffusion mechanism at each network concentration and tracer particle size, we followed the methodology proposed in ref.~\cite{toolbox} and depicted in Fig.\ref{fig:flow_diagram}.

\begin{figure}[ht!]
	\centering
	\includegraphics[scale=0.25]{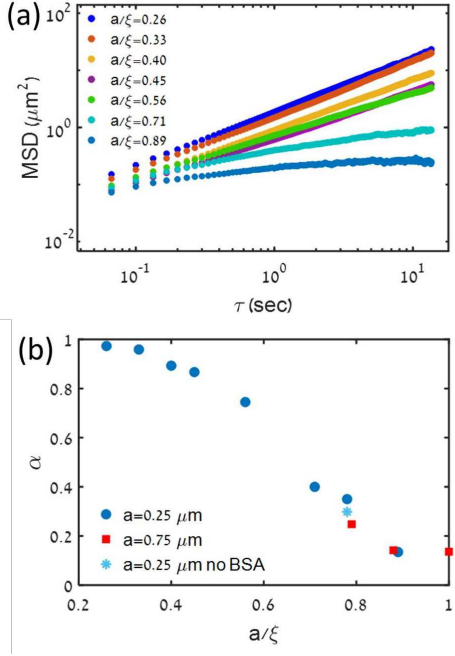}
	\caption{a) MSD as a function of lag time for samples with different a/$\xi$ and a tracer particle radius of a $= 0.25 \mu m$. The motion of the tracer particles becomes more confined as a/$\xi$ increases. b) The diffusion exponent, $\alpha$, as a function of a/$\xi$, measured for samples with two sizes of beads, a$ = 0.25 \mu m$, and $0.75 \mu m$. The pale blue asterisk represents an experiment that was performed using beads that did not undergo surface treatment (see section~\ref{sec:sample}), corresponding to the experiments presented in ref.~\cite{Anomalous2004}. }
	\label{fig:MSD}
\end{figure}

The first test that we performed, in order to classify the diffusion mechanism of the tracer particles within the actin networks, was to extract their typical mean squared displacement averaged over the ensemble of particles EA-MSD,
\begin{equation}
    \langle \Delta \vec{r}^{2}(t) \rangle =\frac{1}{N_t}\sum\limits_{k=1}^{N_t}\left(\vec{r}_k(t)-\vec{r}_k(0)\right)^2,
\end{equation} where $\vec{r}_k(t)$ is the position of the k'th particle at time $t$ and the average is over the ensemble of tracer particles in all similar experiments (Fig.~\ref{fig:MSD}a).

In our analysis, we assumed (and verified) that the system is in a statistically steady state, namely, that its properties do not evolve in time. As shown in Fig.~\ref{fig:MSD}a, for $0.26 < a/\xi < 0.4$, the EA-MSD is close to linear in time, i.e., the particles exhibit normal diffusion. From the slope of the EA-MSD, in this region, we can extract the diffusion coefficient of the tracer particles. Using the Stokes-Einstein relation, we find that the effective viscosity of these networks is on the order of $\sim 4$ times larger than the viscosity of pure water. This is expected due to hydrodynamic interactions with the actin network, \cite{crowd,crowd2,crowd3} and the reduced diffusivity due to the obstacles erected by the actin filaments.

As expected, at larger values of a/$\xi$, the EA-MSD exhibits anomalous sub-diffusion, increasing as a power law for long times. We extract the diffusion power, $\alpha$, from the EA-MSD curves at long times (Fig.~\ref{fig:MSD}b). We see that in the range of $0.26<$a/$\xi<0.89$, the tracer particle diffusion changes significantly. As a/$\xi$ increases, the particles become more confined, and the diffusion power decreases from $\sim 1$ to $\sim 0.13$. These results are in good agreement with previous reports \cite{Anomalous2004}. To verify that the defining factor of the diffusion exponent is the ratio a/$\xi$, rather than the bead radius or the actin concentration, we compared the results obtained by two sets of experiments: the first with tracer particles of radius a$ = 0.25$ $\mu m$ and the second with a $= 0.75$ $\mu m$. Both were performed over a large range of actin concentrations (Fig.~\ref{fig:MSD}b). We found good agreement between the two sets of experiments, confirming that the ratio of the bead size to the mesh size dictates the tracer particles’ diffusion type. As shown in Fig.~\ref{fig:MSD}b, in the range of $0.4 <$ a/$\xi$ $< 0.8$, the diffusion exponent strongly decreases. This sharp decrease accords with the significant change in the individual trajectories of the tracer particles (see Fig.~\ref{fig:trajectoriesVsRatio}) in this range.

\begin{figure}[ht!]
	\centering
	\includegraphics[scale=0.25]{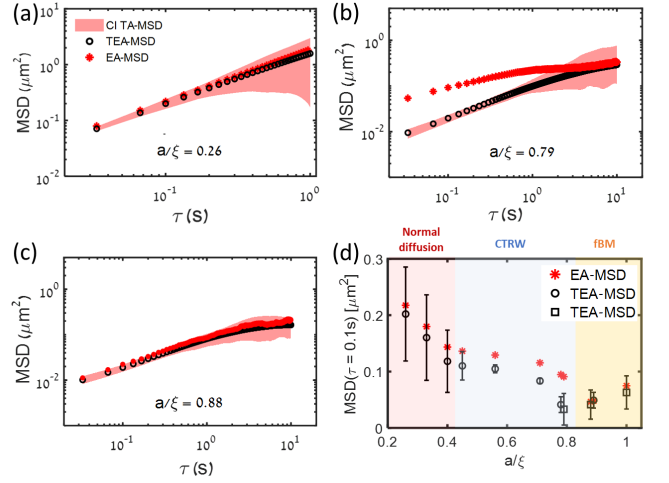}
	\caption{A comparison between the EA-MSD and the TEA-MSD of individual probe particles for (a) a/$\xi= 0.26$, (b) a/$\xi= 0.79$ and (c) a/$\xi = 0.88$. The pink shadow indicates the confidence interval (CI) for the calculated TA-MSD.  (d) The value of the MSD at $\tau$ = 0.1s for actin networks with ratios of probe bead radius to mesh size in the range of $0.26 <$ a/$\xi$ $< 1$. Black circles and squares refer to a $= 0.25$ $\mu m$ and $0.75$ $\mu m$ probe particles, respectively; error bars indicate the 85\% CI for the calculated TEA-MSD. The errors for the EA-MSD are smaller than the symbol size. Background color separation demonstrates the transition between non-ergodic (white background) and ergodic (pink and yellow backgrounds) processes.}
	\label{fig:TEA_EA_TA_MSD}
\end{figure}     

The next classification test, according to the decision tree presented in Fig.~\ref{fig:flow_diagram}, is for ergodicity.
A main difference between the fBm and the CTRW is their ergodicity. It was shown that for a sub-diffusion CTRW (governed by a power-law tailed distribution of the escape time), the process is not ergodic \cite{WEB1,WEB2,WEB3}; namely, the time average and the ensemble average of physical observables are not the same, even in the limit of large ensemble and infinitely long trajectories.
To this end, we compare the ensemble-averaged MSD (EA-MSD) to the time-averaged MSD (TA-MSD) \cite{ergo_test,RWF1}. The TA-MSD is calculated for individual trajectories, and follows the time-dependent correlation function of each particle over time,
\begin{equation}\label{eq:TAMSD}
\overline{\delta^2_{k}}(\tau) = \frac{1}{T_k-\tau}\int\limits_0^{T_k-\tau}\left(\vec{r}_k(t+\tau)-\vec{r}_k(t)\right)^2 dt,
\end{equation}
where $\tau$ is the lag time, $\vec{r}_k(t)$ is the position of the k'th particle at time $t$, $T_k$ is the duration of the k'th trajectory, and $t$ is the integration variable for the temporal averaging.
However, due to experimental limitation, the trajectories of the particles are not long enough to obtain good reliable data at $\tau>0.5$ s (see Fig. S1). Therefore, we calculate the ensemble average of the TA-MSD curves of all particles to obtain the TEA-MSD of all trajectories and compare it to the EA-MSD. The TEA-MSD is defined as:
\begin{equation}\label{eq:TEA-MSD}
    \langle\overline{\delta _k^{2}}(\tau)\rangle\equiv\frac{1}{N_\tau}\sum\limits_{k=1}^{N_\tau}\overline{\delta_k^{2}}(\tau),
\end{equation}
where $N_\tau$ is the number of trajectories spanning a time lag greater than $\tau$. We note that as $\tau$ increases, these two plots should converge, since the TEA-MSD is dominated by the time average at small $\tau$ and by the ensemble average at large $\tau$.
In Fig.~\ref{fig:TEA_EA_TA_MSD}a\textendash c, we show both the EA-MSD and the TEA-MSD of samples with a$/\xi=0.26,0.79,0.88$, corresponding to three regimes of different behaviors. Clearly, only for a$/\xi=0.79$ is the diffusion non-ergodic. As expected, the difference is more pronounced for smaller lag times where the time average in the TEA-MSD curves dominates the ensemble average. Therefore, we choose to evaluate ergodicity from the difference in the value of the EA-MSD and TEA-MSD at $\tau=0.1s$ for different a$/\xi$ values in Fig.~\ref{fig:TEA_EA_TA_MSD}d. We identify three distinct regimes based on the flow diagram in Fig.~\ref{fig:flow_diagram}: normal-like diffusion at a$/\xi<0.4$, CTRW-like diffusion at $0.4<$a$/\xi<0.8$, and an ergodic sub-diffusion process (fBm or another ergodic mechanism) at $0.8<$a$/\xi<1$.

\begin{figure}[ht!]
	\centering
	\includegraphics[scale=0.2]{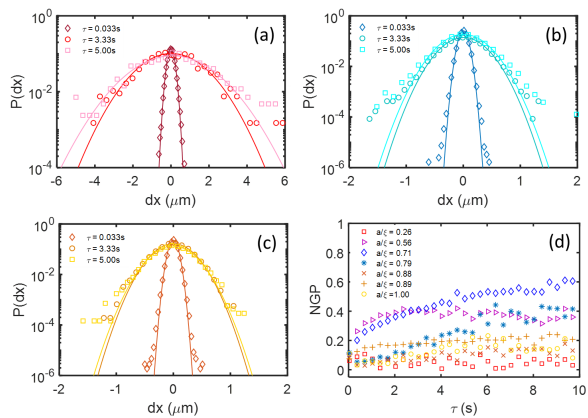}
	\caption{PDF for different lag times for the ratio of probe bead radius to mesh size of (a) a/$\xi = 0.26$, (b) a/$\xi = 0.79$, and (c) a/$\xi = 0.88$. Black lines indicate a Gaussian fit to the short and long lag times. (d) NGP as a function of $\tau$ for actin networks with ratios of probe bead radius to mesh size within the range of $0.26 <$a/$\xi< 1$.}
	\label{fig:NGP}
\end{figure}     

Finally, in order to decide whether the tracer particles diffuse in actin networks in the range of a/$\xi>0.8$ according to fBm or another mechanism, we extracted the step size probability distribution function (PDF) from their trajectories at different lag times and fitted it to a Gaussian curve, which is the final test suggested in Fig.~\ref{fig:flow_diagram} and ref. \cite{toolbox,RWF1}.
In Fig.~\ref{fig:NGP}a\textendash c, we show typical PDFs for samples in the three regimes. We then proceeded to evaluate the Gaussianity of the PDFs for all samples at all lag times using the non-Gaussian parameter (NGP), which is given by:
\begin{equation}
\text{NGP}(\tau) = \dfrac{\overline{\langle \Delta {x}(\tau)^{4} \rangle}}{3\overline{\langle \Delta {x}(\tau)^{2} \rangle}^2} - 1,
\label{eq:NGP}
\end{equation}
where $\Delta {x}(\tau)$ is the projected displacement on one axis ($\hat{x}$ or $\hat{y}$) within a time window of $\tau$ averaged over both time and ensemble. The NGP represents the deviation of the kurtosis from the Gaussian value, and approaches zero for Gaussian distributions, \cite{rahman1965}.
The lag time dependence of the NGP, for different values of a$/\xi$, is depicted in Fig.~\ref{fig:NGP}d.
For a$/\xi=0.26$, we find a close to Gaussian PDF at all lag times; this further confirms that the diffusion in this regime is normal with no considerable effect of confinement by the network. For the second regime, where $0.56\leq$a$/\xi\leq0.79$, we find that the PDF is not Gaussian at all time scales. The deviation from a Gaussian PDF, which is expected due to the considerable confinement by the actin network, cannot uniquely explain the mechanism of the dynamics. For the last measured regime, where $0.8<$a$/\xi<1$, we find that the deviation from Gaussian PDF is smaller (than the deviation found for the second regime) for long time lags. This is in agreement with the fact that for the large a$/\xi$ ratios, the tracer particles only visited a few cages.    
\begin{figure}[ht!]
	\centering
	\includegraphics[scale=0.22]{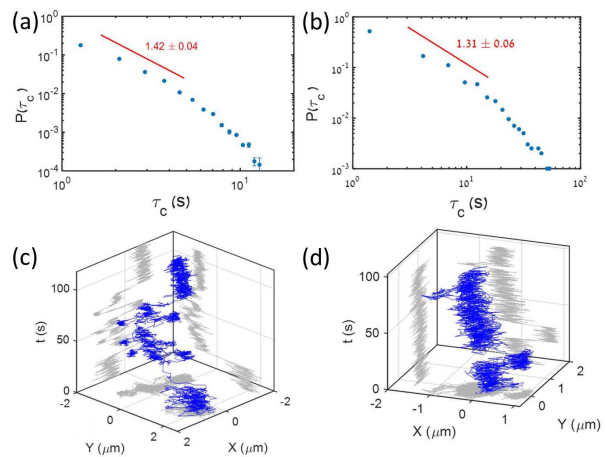}
	\caption{Histogram of caging times for diffusing particles in an entangled F-actin network with particle radius to mesh size ratios of (a) a/$\xi= 0.71$ and (b) a/$\xi = 0.78$. The red line indicates a power-law fit, $P(\tau_c) \sim \tau_c^{-\nu}$ , with $\nu = 1.42 \pm 0.04$ and $1.31 \pm 0.06$, respectively. (c,d) Representative trajectories of tracer particles projected on the x-y plane as a function of time for samples with a/$\xi = 0.71$ and $0.78$, respectively.}
	\label{fig:P_tau_ratio078_a_025um}
\end{figure}

As already demonstrated in Fig.~\ref{fig:trajectoriesVsRatio}c\textendash e, at intermediate values of a/$\xi$, particles jump randomly between different micro-environments in which they are constrained. This can be seen more convincingly when plotting the trajectory in 3D with time as the $z$ axis (see Fig.~\ref{fig:P_tau_ratio078_a_025um}a,b). Such dynamics can be characterized by the probability distribution of caging times $P(\tau_c)$ within the local micro-environments, which according to CTRW theory, should be scaled as a power law with the caging time, i.e., $P(\tau_c) \sim \tau_c^{-\nu}$. We extracted $P(\tau_c)$ directly from the trajectories by identifying the onset of caging and the escape. We then accumulated data from all the trajectories in the sample and plotted them as a function of $\tau_c$. For example, we show the caging time distribution for a$/\xi=0.71$ and a$/\xi=0.78$ in Fig.~\ref{fig:P_tau_ratio078_a_025um}c,d. In these plots, we observe that at larger values of $\tau_c$, the power law decreases, possibly due to the lack of sufficient statistics (in accordance with \cite{Anomalous2004}). We, therefore, extract the power law from the slope at small $\tau_c$.

\begin{figure}[ht!]
	\centering
	\includegraphics[scale=0.24]{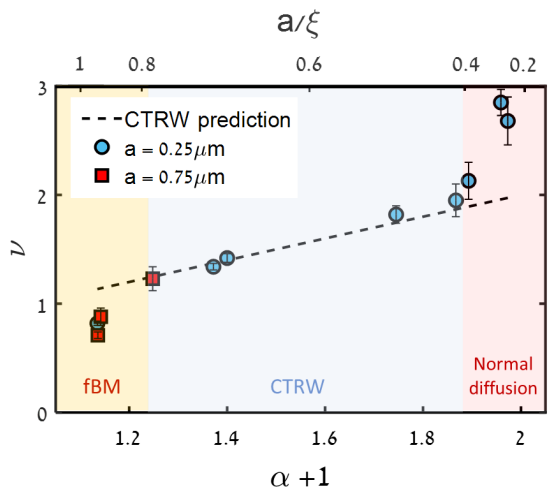}
	\caption{Power-law fit value ($\nu$), extracted from the relation $P(\tau_c) \sim \tau_c^{-\nu}$ for ratios of probe bead radius to the actin network mesh size in the range of $0.26 <$ a/$\xi< 1$, versus the observed diffusion exponent plus one ($\alpha+1$).  For CTRW, we expect $\nu=\alpha+1$. Only within the range $0.4 <$ a/$\xi< 0.8$ is there a good agreement with CTRW theory, while for lower and larger ratios, there are significant deviations that indicate that the diffusion is dominated by other mechanisms. Based on the results that we obtained previously, we can deduce that these diffusion mechanisms are fBm for a/$\xi>0.8$ and free diffusion for $0.26 <$ a/$\xi<0.4$.}
	\label{fig:CTRW}
\end{figure}

The CTRW model describes anomalous diffusion arising from random walks with discrete steps of constant velocity, separated by pauses of random duration. When the direction of the random steps is chosen symmetrically and $1 < \nu < 2$, CTRW predicts that the MSD should be sub-diffusive, scaling asymptotically as $\alpha = \nu -1 $ for uncorrelated jumps \cite{Shlesinger1974,WEEKS1996}.
In order to test the validity of the relation between the diffusion exponent and the caging power law, we calculated $P(\tau_c)$ for all a/$\xi$ values, and extracted the power law, $\nu$. In Fig.~\ref{fig:CTRW}, we plot $\nu$ as a function of $\alpha+1$. The three diffusion regimes are clearly observed in this analysis, confirming that for $0.4<$a$/\xi<0.8$, particles diffuse in a CTRW-like process. Fig.~\ref{fig:CTRW} shows a summary of all the diffusion mechanisms that were revealed for the tracer particles in the entangled actin gel for the whole range of $0.26 <$ a/$\xi< 1$ studied here. It is important to note that for the CTRW-like regime, we tested the data for an aging signature, expected for a CTRW \cite{Schulz2013}. However, we could not identify any signatures of aging over the time scale of the measurements.

\begin{figure}[ht!]
	\centering
	\includegraphics[scale=0.3]{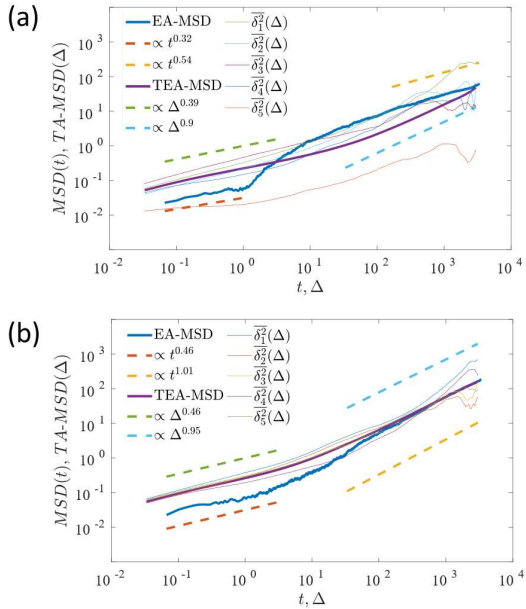}
	\caption{The simulated ensemble averaged square displacement (EA-MSD), ensemble average of time averaged square displacement (TEA-MSD), and time averaged square displacements (TA-MSD). (a) Caged fBm with power-law distributed caging times and (b) caged fBm with exponentially distributed caging times. For both cases, there is a clear transition between the short time (within the cage) and the long time (including hopping between cages) dynamics. The solid lines represent the EA-MSD and the TEA-MSD, the dashed lines represent the fitted power-law curves, and the thin solid lines represent the TA-MSD of typical trajectories.}
	\label{fig:sim}
\end{figure}

We propose caged fBm as a conceptual model for the tracer dynamics in actin networks observed here. The viscoelastic response of the actin network to perturbations induced by the thermal motion of the tracer particle is long lived and has memory. This response will induce a fBm of the tracer particle when it is large enough to have almost continuous contact with the filaments (i.e., when $a/\xi>0.5$). However, the filaments also confine the tracer’s motion. Long range motion is possible only when the tracer hops between cages formed by the actin filaments. For this type of dynamics, one expects a transition from a fBm, for short times while the tracer is mostly within a cage, to a different longer time dynamics that is dictated by the distribution of the caging times (or escape times). For a caging time distribution with finite moments, the long time asymptote is a normal diffusion (the mean escape time of a particle undergoing fBm is finite); for a power-law distribution of caging times with a diverging first moment, the long time dynamics is characterized by a CTRW-induced anomalous diffusion. We found that the cage size distribution is relatively narrow and that there is no correlation between the cage size and the escape time from the cage. These two properties hint that the escape and the caging depend on the large bending fluctuations of actin filaments that release the trapped tracer. To test our proposed model, we simulated two variants of the caged fBm model, with either a power-law or an exponential distribution of the caging times. For each distribution, we simulated 1000 trajectories. The results are presented in Fig. \ref{fig:sim}. Based on experimental evidence, we used a gamma distribution of cage sizes. This was the source of the different TA-MSD in the case of the exponentially distributed caging time. For the power-law distributed caging time, the TA-MSD is a random variable and the TEA-MSD is not equal to the EA-MSD due to the ergodicity breaking. In both cases, the short time behavior is dictated by the fBm, while the long time behavior is different between the two cases, as described above.
The measured dynamics appear to be similar to the results for a caged fBm with a power-law distribution of caging times. The supporting characteristics are the fact that the spread of the TA-MSDs does not seem to grow with time (before the number of trajectories becomes very small), and the fact that the EA-MSD is larger than the TEA-MSD at intermediate times. However, for the measured data, there are limitations (there is a smaller number of long trajectories, and the statistics at long lag times is dominated by the ensemble average) that do not exist for the simulated trajectories. Therefore, the results are not expected to fully coincide.

\section{Concluding remarks}

Our results suggest that the diffusion characteristics of the tracer particles are considerably affected by the ratio a/$\xi$ in the regime where the sizes of the tracer particle and the network mesh are comparable (a $\sim \xi$). We observed that the surprising CTRW-like motion of tracer particles in an in-vitro polymerized actin network belongs to one of three diffusion mechanisms that are seen in these networks. For small ratios, the particles perform normal diffusion through the network pores, at intermediate ratio values they perform CTRW-like motion, and at large ratios, they perform fBm, as was originally expected. The physical processes leading to the diffusion at high and low ratio values are well understood; however, the physical reasons for the CTRW-like motion at intermediate values of a$/\xi$ are not clear. Clearly, the tracer particles in these conditions hop between adjacent compartments within the network (Fig.~\ref{fig:P_tau_ratio078_a_025um}c,d). However, the physical reason for a power-law distribution of escape times from a compartment (at least at short times, Fig.~\ref{fig:P_tau_ratio078_a_025um}a,b) is yet to be determined.

It is tempting to try and describe the dynamics as a sequence of escapes from cages. For small tracer particles, one may naively assume that dynamics within the compartment is simply a normal diffusion (with a probably reduced diffusivity, as was indeed observed, due to the hydrodynamic effects imposed by the network). For normal diffusion, the mean escape time from a cage is finite and scales as the square of the cage size divided by the diffusion coefficient. Therefore, assuming that the cage does not change much during the measurement time, one would simply expect a transition between diffusion within the cage at short times and diffusion between cages at long times (note that if the caging/trapping is not significant, the diffusivity within the cage and between cages is expected to be very similar). Moreover, the probability distribution of escape times (or residence times within the cage) is expected to have an exponentially decaying tail. Note that this type of dynamics resembles the observed dynamics for a small ratio of a$/\xi$.

For tracer particles of a size similar to the network mesh size, the interaction of the particle with the network filaments is expected to induce a fBm dynamics. For a particle exhibiting fBm, the mean escape time from a cage scales like the size of the cage to the power $2/\alpha$ ($0\leq\alpha\leq 1$ is the diffusion exponent) \cite{Guggenberger_2019}. The probability density function of the escape time is expected to have a stretched exponent decaying tail. Nevertheless, all the moments of the escape time are finite, and the dynamics over long time scales are expected to exhibit normal diffusion.
There is no evidence for a broad distribution of cage sizes that may result in a CTRW. Therefore, the origin of the CTRW-like dynamics is likely to be the interaction between the tracer particle and the actin filaments, which triggers the large bending of these filaments, thereby enabling the escape from the cage. It is also important to note that for a CTRW, the TA-MSD is expected to be linear in the time lag (unlike the EA-MSD). However, our results show that it is not. Moreover, we did not find any evidence for aging, as would be expected for a CTRW.
The power-law waiting time PDF and its correspondence with the diffusion exponent support the CTRW dynamics. A possible explanation for the discrepancies between CTRW and the observed dynamics may be the fact that our trajectories span a relatively small number of cages, and therefore, the dynamics is not expected to converge to the asymptotic limit of CTRW.

Our simulation results show that anomalous diffusion may be exhibited by particles undergoing caged fBm, regardless of the distribution of the caging times. The TA-MSD of individual tracers may be different due to the heterogeneity of the cage sizes (even when the distribution is narrow) and the finite duration of the measurement. However, in order to tightly relate the model to the experiment, one has to understand the mechanism of escaping from the actin cage.

In our analysis, we analyzed trajectories of approximately equal and relatively long length (2000 frames$\approx$67s). This was done to allow for an equal and sufficient amount of statistics in each analysis. Long trajectories are commonly required to compare measurements to the asymptotic behavior described in the theoretical model. In many experimental scenarios, long trajectories are unavailable due to technical issues, such as the finite fluorescence lifetime of tagged molecules, or due to fast or changing dynamics of the tracer particles. Therefore, there is a need to develop alternative classification schemes for diffusion processes that can be obtained from relatively short trajectories and can be tied to the underlying physics. Here, we used the step-size distribution that could be extracted from a large ensemble on shorter trajectories. Similarly, to some extent, the escape time distribution, up to a given cutoff, can be obtained from such an ensemble. This is a first step toward such a new classification.

One important implication of our results is related to the characterization of biological systems and processes using single molecule tracking. Namely, the type of motion of a tracer particle, or a molecule, in a complex material depends dramatically on the relation between its size and the typical length scales of the material. One approach to overcome this challenge can be to use tracers of different sizes. An alternative powerful tool would be to use the correlations in the motion of the tracer particles, as is done in two-point microrheology \cite{Crocker2000}. The correlated diffusion carries structural information \cite{gardel2006, Oppenheimer2011,Adar2014_2} even when applied to systems out of thermal equilibrium \cite{Sonn_Segev_2017,Chein2019}.

\section*{Acknowledgments}
The authors are grateful to Anne Bernheim-Groswasser for the numerous discussions and for providing actin. ML and YR thank the Center for Light Matter Interaction in Tel-Aviv University for its support. This work was partially supported  by the US-Israel Binational Science Foundation grant 2014314 and by the ISF (Israel Science Foundation) grant 988/17.

\end{document}